\documentclass[5p,superscriptaddress]{elsarticle}

\usepackage{lineno,hyperref}
\usepackage{graphicx} 
\usepackage{textcomp}
\usepackage{amsmath}
\modulolinenumbers[5]

\journal{Scripta Materialia}

\biboptions{square,comma,numbers,sort&compress}

\begin{document}

\title{A comparison of in- and ex situ generated shear bands in metallic glass by transmission electron microscopy}

\author[IMP]{Harald R\"osner\corref{correspondingauthor}}

\ead{rosner@uni-muenster.de}

\author[KIT,MNF,DA]{Christian K\"ubel}

\author[IMP]{Stefan Ostendorp}

\author[IMP]{Gerhard Wilde}

\cortext[correspondingauthor]{Corresponding author}

\address[IMP]{Institut f\"ur Materialphysik, Westf\"alische Wilhelms-Universit\"at M\"unster, Wilhelm-Klemm-Str. 10, 48149 M\"unster, Germany}

\address[KIT]{Karlsruhe Institute of Technology (KIT), Institute of Nanotechnology (INT), Hermann-von-Helmholtz-Platz 1, D-76344 Eggenstein-Leopoldshafen, Germany}

\address[MNF]{Karlsruhe Nano Micro Facility, Karlsruhe Institute of Technology, D-76344 Eggenstein-Leopoldshafen, Germany}

\address[DA]{Material- and Geowissenschaften, TU Darmstadt, Alarich-Weiss-Str. 2, 64287 Darmstadt, Germany}

\date{\today}

\begin{abstract}
Shear bands originating from in situ tensile tests of Al$_{88}$Y$_{7}$Fe$_{5}$ melt-spun ribbons conducted in a transmission electron microscope are compared with ones which had formed ex situ during cold rolling. During in situ straining, the observations of a spearhead-like shear front, a meniscus-like foil thickness reduction and no apparent shear steps to accommodate strain suggest shear band initiation by a rejuvenating shear front followed by shearing along the already softened paths. This leads to necking and subsequent failure under the reduced constraint of a 2D geometry in the thin foil and thus explains the observed lack of ductility under tension. In contrast, shear bands formed during cold rolling display distinct alternating density changes and shear off-sets. An explanation for this difference may be that in situ shear bands rip before such features could develop. Moreover, both in and ex situ experiments suggest that initiation, propagation and arrest of shear bands occur during different stages. 
\end{abstract}


\maketitle

Deformation processes in metallic glasses are quite different from those in crystalline materials due to the absence of a periodic lattice. Deformation tests on metallic glasses well below the glass transition temperature using deformation rates less than 10$^{-2}$ have shown that the plastic flow is confined to narrow regions called shear bands when the applied strain exceeds the elastic range\cite{pampillo1972localized,spaepen1977microscopic,schuh2007mechanical,greer2013shear,maass2015shear,hufnagel2016deformation}. Thus, shear bands are a precursor to material failure 
and typically have widths of  5 - 20\,$nm$\cite{donovan1981structure,li2002nanometre,zhang2006thickness,shao2013two,rosner2014density,hieronymus2017shear}. However, larger widths in the range of 100 - 200\,$nm$ have also been reported \cite{pauly2009crack,liu2017shear,maass2020beyond,liu2021strain}. 

\noindent It is commonly believed that the shear band core is associated with a structural change due to shear dilatation, implying a volume increase and thus a change in density\cite{donovan1981structure,li2002nanometre,li2002characterization,jiang2003effect,jiang2005mechanical,hajlaoui2006shear,zhang2006thickness,schuh2007mechanical,ishii2008relaxation,guan2010stress,lechner2010vacancy,klaumunzer2011probing,miller2011detecting,pan2011softening,shao2013high,greer2013shear,maass2015shear,liu2017shear,liu2019shear,maass2020beyond,liu2021strain}. Therefore, shear band cores are softer\cite{bei2006softening,pan2011softening} than the surrounding matrix allowing the applied shear strains to accommodate via slip. An important issue is thus the quantification of free volume\cite{klaumunzer2011probing} or density inside shear bands\cite{rosner2014density}. 

\noindent While the interpretation of contrast changes in transmission electron microscopy (TEM) is often complicated\cite{hirsch1977electron,edington1977practical,williams1996transmission}, local contrast changes within shear bands of an Al$_{88}$Y$_{7}$Fe$_{5}$ metallic glass have been successfully determined as density changes using high angle annular dark field scanning transmission electron microscopy (HAADF-STEM) together with foil thickness measurements\cite{rosner2014density,schmidt2015quantitative,grove2021plasmon}. Further, ex situ experiments on Zr- and Pd-based metallic glasses as well as simulations showed that often high and low density regions alternate along the propagation direction of the shear bands\cite{hieronymus2017shear,hassani2019probing,hilke2019influence,Davani2019,liu2021strain}. Additional measurements using atom probe tomography confirmed that the local contrast changes within shear bands are correlated with density changes\cite{balachandran2019elemental,chellali2020deformation,mu2021unveiling}. Hence, both densification and dilation may occur locally within the same shear bands as a response to plastic deformation in the bulk. Moreover, it has been found that there are more low density regions than high density regions\cite{schmidt2015quantitative}, which explains why macrosopic shear band measurements show a volume increase or density decrease\cite{lechner2010vacancy,klaumunzer2011probing,shao2013two}. 

\noindent Monolithic metallic glasses lack ductility, especially under tension where immediate catastrophic failure predominates after reaching the yield point\cite{ashby2006metallic}. However, when hydrostatic pressure is present such as in cold rolling, bending or compression testing, monolithic metallic glasses do exhibit ductility\cite{bei2006softening,schuh2007mechanical,eckert2007mechanical,scudino2011ductile,okulov2015flash,nollmann2016impact,scudino2018ductile,bian2019controlling,peng2019deformation,kosiba2019modulating}. Different models have been proposed to explain shear band formation in metallic glasses, on which the apparent ductility relies\cite{packard2007initiation,schuh2007mechanical,greer2013shear,maass2015shear}. We focus here on the following three scenarios: The first model suggests that local structural transitions known as shear transformation zones (STZ)\cite{argon1979plastic} are activated in the entire plane of the highest resolved shear stress. They subsequently align in an autocatalytic percolation process forming a shear band\cite{packard2007initiation,maass2015shear,csopu2017atomic}. This process requires the coordinated reshuffling of many atoms in concert. Moreover, this scenario displays homogeneous shear band nucleation as it nucleates from structural fluctuations rather than from flaws serving as heterogeneous nucleation sites. The second model\cite{shimizu2006yield} suggests that defects in the cast material such as voids and surface notches serve as stress concentrators when the external load is applied. Shear bands may preferentially initiate from such sites by activating a group of STZs forming an embryonic shear band, which after reaching a critical size of about $100$\,$nm$ in length, would propagate and develop into a mature shear band displaying a 'spearhead-like hierarchy' of the shear front, which is thought to depict different zones representing different glass states, i.e. aged, rejuvenated, glue and liquid\cite{shimizu2006yield}. Thus, this model requires frictional heat dissipation from a fast propagating shear front to cause a large temperature increase such that the glass transition is reached/passed. The third model considers shear band formation as involving different stages: the first stage describes heterogeneous nucleation by structural rejuvenation in which a softened path is formed by the activation of individual STZs. During this stage the shear strain as well as the temperature rise is small. The second stage involves shearing along the already softened path leading to local heating and shear band thickening\cite{cao2009structural,greer2013shear}. 

\noindent In this letter, we compare in situ generated shear bands formed under tension in the TEM and ex situ shear bands formed during cold rolling. In situ experiments provide additional information about shear band mechanisms in combination with size-effects\cite{matthews2008electron,volkert2008effect,de2009advances,jang2010transition,chen2010effects,greer2011plasticity,wilde2011nanocrystallization}.

Fully amorphous ribbons of Al$_{88}$Y$_{7}$Fe$_{5}$ (composition in atomic percent) with an average thickness of $40\,\mu m$ were produced by melt spinning. For more details see Ref.\cite{bokeloh2010primary}. Shear bands were produced by in situ straining of the ribbons at ambient temperature using a single-tilt tensile stage (Gatan Model 672)\cite{rosner2009situ} and alternatively, ex situ by cold rolling the ribbons to a thickness reduction of 10\,$\%$. TEM specimens were prepared by twinjet electro-polishing (Tenupol-5, Struers) using HNO$_3$:CH$_3$OH in a ratio 1:2 at -22 $\,^\circ$C applying voltages of about -10.5 $\,V$. The in situ TEM study was performed using a FEI Titan 80-300 image-corrected transmission electron microscope equipped with a post-column energy filter (Tridiem 863 Gatan Imaging Filter) operated at 300\,$kV$ in STEM mode. HAADF images and electron-energy loss (EEL) spectra were collected with a HAADF detector (Fischione model 3000) and a slow scan CCD camera (Gatan US 1000) using the following parameters: camera length of 102$\,mm$, convergence semi-angle $\alpha$ of 9.5$\,\,mrad$, collection semi-angle $\beta$ of 3$\,\,mrad$ with $\alpha > \beta \gg \theta_{E}$, an entrance aperture of 2$\,mm$, an energy dispersion of 0.2 $eV$/channel, an acquisition time of 400  $\,ms$, and a nominal spot size of 0.5$\,nm$. The characterization of the cold-rolled samples was performed with a Zeiss Libra 200FE TEM (Fig.~\ref{FIG3}) and a Thermo Fisher Scientific Themis 300 G3 transmission electron microscope (TEM) (Fig.~\ref{FIG4}) both operated in STEM mode. An atomic force microscope (AFM, Park XE 100) operated in non-contact-AFM mode was used to measure the topography of the sheared zones after deformation. 

\noindent The HAADF-STEM signal (electrons collected by the HAADF detector between 60 $mrad$ and 200 $mrad$) can be used to gain information about the density/volume change\cite{schmidt2015quantitative}. The cross-section for HAADF scattering approaches the (un)-screened Rutherford cross-section which is in the range of $Z^{1.7-2}$\cite{pennycook2002structure}.

\noindent An exponential decrease in transmission for the STEM signal with increasing mass thickness $x = \rho$ $\cdot \,t$ has been found for amorphous specimens\cite{reimer1976recording,kohl2008transmission}. The dark-field intensity $I/I_0$ can be formulated in the form:

\begin{equation}
\frac{I}{I_0} = \left[ 1 - exp\left( -\frac{\rho \cdot t}{x_k} \right) \right] 
\end{equation}

where $\rho$ is the density, $t$ is the foil thickness and $x_{k}$ is the contrast thickness. For a constant contrast thickness $x_{k}$ and a small argument\cite{rosner2014density}, the HAADF-STEM signal $I/I_0$ scales with the mass thickness $(\rho \cdot \,t )$:
\begin{equation}
\frac{I}{I_0} \propto \rho \cdot t 
\end{equation}

\noindent To determine the density one needs to measure the corresponding local foil thickness t\cite{malis1988eels}. This can be achieved by EEL spectroscopy (EELS) using the information of the low-loss region. The refractive index-corrected Kramers-Kronig sum rule according to Iakoubovskii et al.\cite{iakoubovskii2008thickness} was used to calculate the corresponding foil thickness profiles, as it is thought to provide more accurate values for the foil thickness in amorphous materials than the log-ratio method\cite{malis1988eels}. It analyses the single scattering distribution $S(E)$, which is obtained from the EEL spectrum. Plural scattering was removed before the thickness computation from the EEL spectra using the Fourier-Log method\cite{egerton1985fourier}. A systematical error for the absolute foil thickness is in the range of $\pm 20\,\%$\cite{malis1988eels,egerton2011electron}. Finally, a density change can be derived after Eq.\,(2) by measuring the contrast signal (intensity) together with the corresponding foil thickness. To avoid contamination during measurements the samples were plasma-cleaned in pure Ar prior to analysis.

\begin{figure}[h!]
	\centering
	\includegraphics[width=\columnwidth]{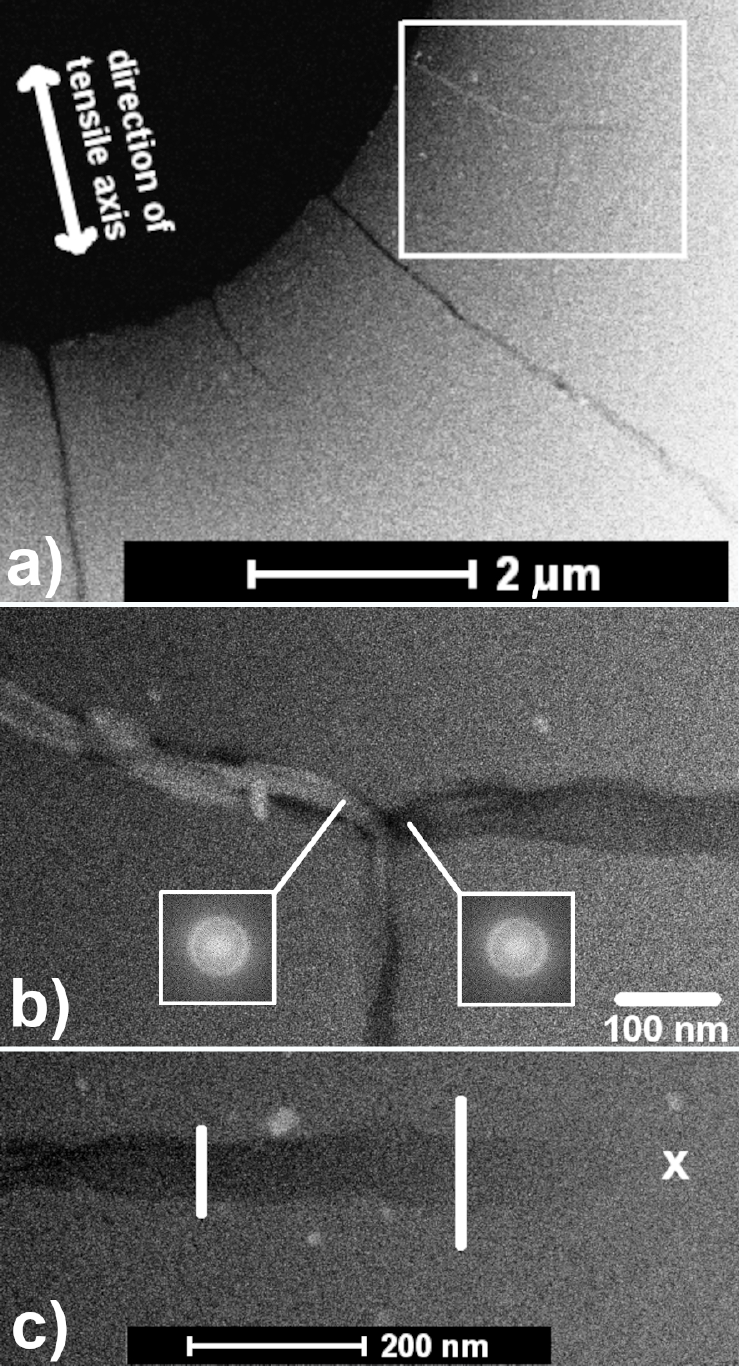}
	\caption{(a) HAADF-STEM image showing an overview of shear bands generated during an in situ TEM tensile test of an Al$_{88}$Y$_{7}$Fe$_{5}$ melt-spun ribbon. The framed area shows a shear band with a branch. (b) Enlarged view showing details of the intersection. The left side shows amorphous material clinging to the shear band. Fourier transforms performed from a high-resolution micrograph (see Supplementary Material, Fig. S1) are shown as insets revealing the amorphous nature of both shear band parts. (c) Enlarged view of the tip of the horizontally propagating part of the shear band. The white lines indicate the position of the line profiles performed from bottom to top across the shear band. The position marked X indicates the shear band tip revealing a spearhead-like shape.}
	\label{FIG1}
\end{figure}

\begin{figure}[h!]
	\centering
	\includegraphics[width=\columnwidth]{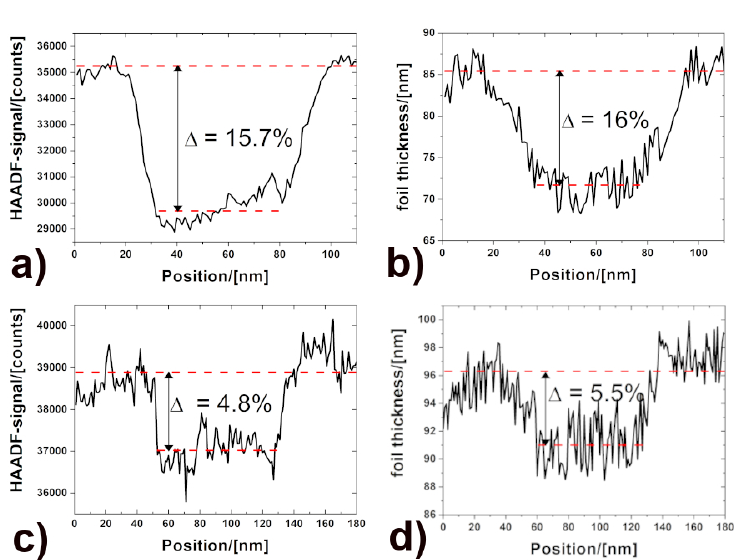}
	\caption{Corresponding line profiles as indicated in Fig.~\ref{FIG1}c. (a) HAADF-STEM profile across the shear band corresponding to the shorter line showing a meniscus-like reduction of $(15.7 \pm 1.9)\,\%$. (b) Corresponding thickness profile showing a meniscus-like reduction of $(16 \pm 2.8)\,\%$. (c) HAADF-STEM profile across the shear band corresponding to the longer line showing a drop of $(4.8 \pm 1.9)\,\%$. (d) Corresponding foil thickness profile showing a reduction of $(5.5 \pm 2)\,\%)$.}
	\label{FIG2}
\end{figure}

\begin{figure*}[htbp]
	\includegraphics[width=\linewidth]{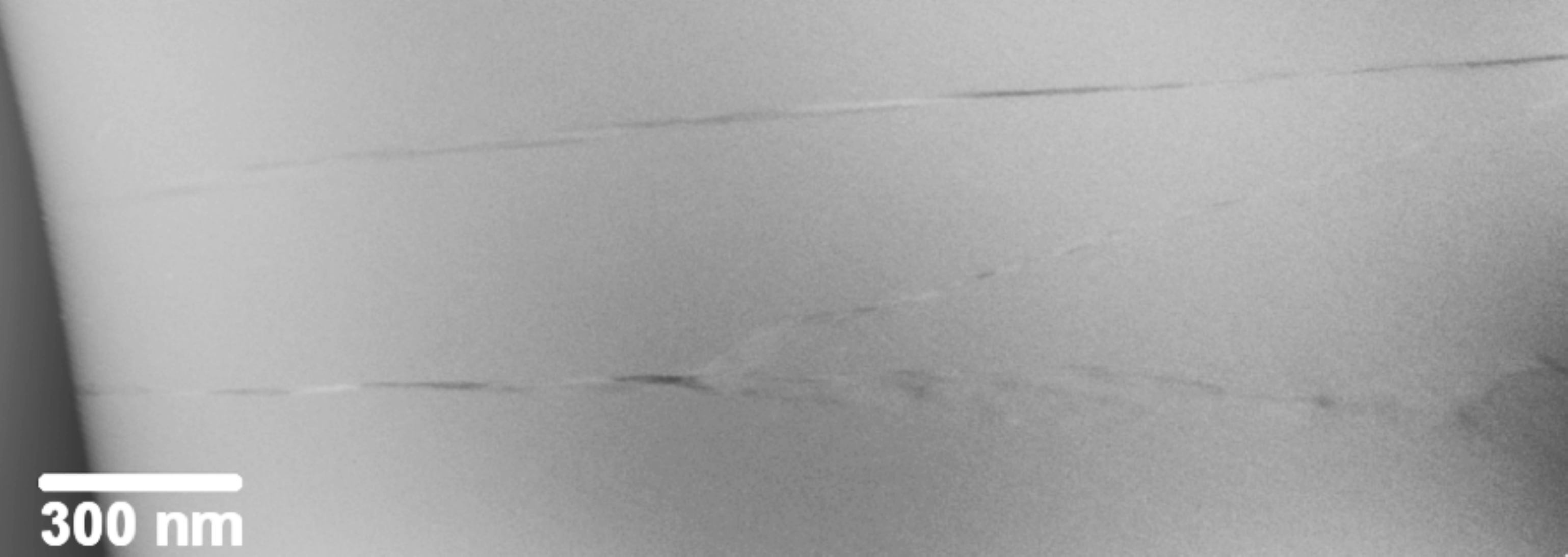}
	\caption{HAADF-STEM image showing an overview of two shear bands produced during cold rolling Al$_{88}$Y$_{7}$Fe$_{5}$ melt-spun ribbons. Note that the lower one shows a bifurcation further along its propagation.}
	\label{FIG3}
\end{figure*}

\begin{figure*}[htbp]
	\includegraphics[width=\textwidth]{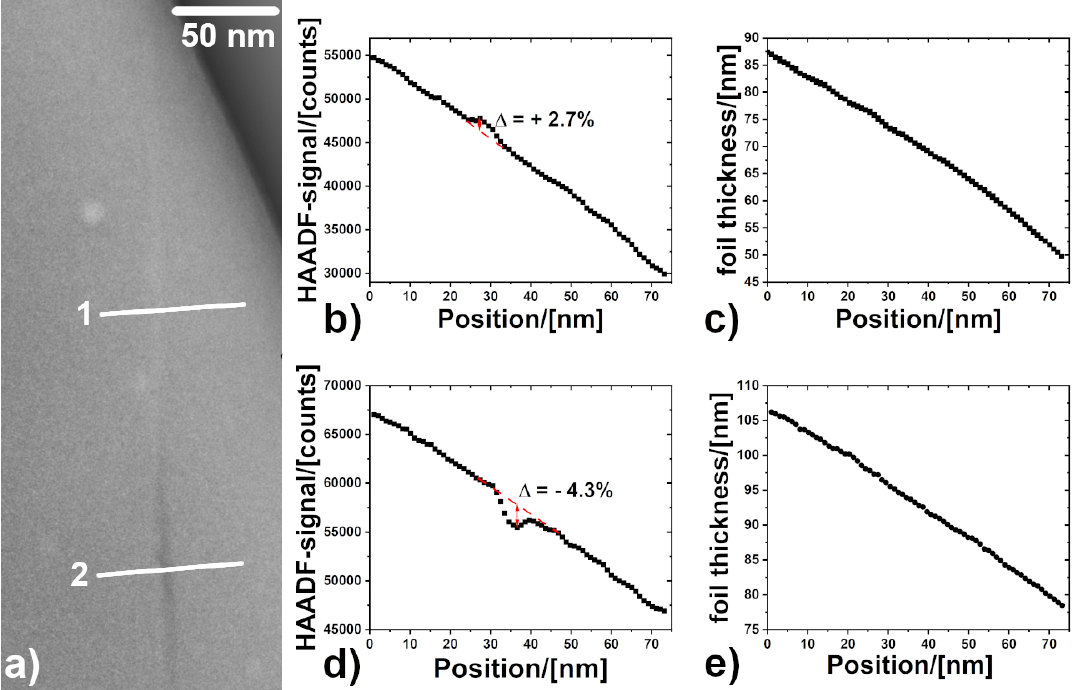}
	\caption{(a) HAADF-STEM image showing details of the alternating contrast change from another shear band produced by cold rolling (ex situ). The white lines indicate the position of the line profiles performed from left to right across the vertical shear band. (b) HAADF-STEM profile corresponding to line 1 showing an upward bump of $(+2.7 \pm 0.4)\,\%$ in the signal for the bright shear band segment. (c) Corresponding thickness profile showing a continuous decreasing signal due to the wedge-shape of the TEM sample. (d) HAADF-STEM profile corresponding to line 2 showing a downward bump of $(-4.3 \pm 0.3)\,\%$ in the signal for the dark shear band segment. (e) Corresponding foil thickness profile showing a continuous decreasing signal due to the wedge-shape of the TEM sample.}
	\label{FIG4}
\end{figure*}

During the in situ elongation of the sample, the edge of the electron transparent area was observed at low magnification in STEM mode. Scanning at low magnification reduces irradition effects\cite{mavckovic2016situ} such as nanocrystallization\cite{wilde2011nanocrystallization} and provides a larger field of view. When shear bands occurred, the deformation was immediately stopped. Fig.~\ref{FIG1}a displays a HAADF-STEM image showing shear bands triggered by the elongation of the foil. Most of them are torn open. The shear bands do not show shear steps at the edge of the thin film. The framed area shows a shear band with a branch. The first part of this shear band appears bright due to amorphous material clinging to the shear band (Fig.~\ref{FIG1}b). Fig.~\ref{FIG1}c shows an enlarged view of the tip of the horizontally propagating part of the shear band and confirms the spearhead-like shear front proposed in Ref.\cite{shimizu2006yield}. The visible width of the shear band varies around 75\,\,$nm$ until it tapers towards the tip. The cross marks the shear band tip. The white lines indicate the position of the line scans performed in nanometer steps across the shear band. The HAADF-STEM signals corrected for the vacuum off-set are shown in Figs.~\ref{FIG2}a and Fig.~\ref{FIG2}c. The signals decrease from the matrix level by $(15.7 \pm 1.9)\,\%$ and $(4.8 \pm 1.9)\,\%$, respectively. The corresponding thickness profiles across the shear band (Fig.~\ref{FIG1}c) are plotted in Fig.~\ref{FIG2}b and Fig.~\ref{FIG2}d showing drops of $(16 \pm 2.8)\,\%$ and $(5.5 \pm 2)\,\%)$. All shear band profiles shown in Fig.~\ref{FIG2} indicate a meniscus-like relief. Profiles obtained post deformation by non-contact AFM measurements from the same sample revealed a similar topology (see Supplementary Material, Fig. S2). Since both intensity and thickness profiles in Fig.~\ref{FIG2} show similar drops within the error margin, the contrast change can be attributed according to Eq.\,(2) to thickness reductions rather than to density changes\cite{donovan1981structure,de2009advances,liu2017shear}.

\noindent For comparison, Fig.~\ref{FIG3} shows shear bands formed during cold-rolling imaged under edge-on or near edge-on imaging conditions displaying alternating contrasts. The lower shear band is aligned parallel to the electron beam (edge-on condition) whereas the upper one is slightly tilted revealing a misalignment between the bright and dark shear band segments. Each contrast reversal is accompanied by a slight deflection of the shear band from the main propagation direction. The shear band width is here about

\noindent 4\,\,$nm$ and so, is much smaller than the in situ generated shear bands shown in Fig.~\ref{FIG1}. Moreover, a bifurcation can be noticed further along the lower shear band (Fig.~\ref{FIG3}) displaying a weaker contrast and a broader appearance in the projection as a result of the shear band rotation\cite{csopu2017atomic,csopu2020atomic}.

\noindent Fig.~\ref{FIG4} shows details of another shear band produced by cold-rolling where the contrast switches from bright to dark. The HAADF-STEM signal and the foil thickness were measured at the indicated positions for the different shear band segments. Both foil thickness profiles (Fig.~\ref{FIG4}c,e) show a continuous decreasing signal as expected for the wedge-shape of the TEM sample. No meniscus occurs across the shear band. However, the HAADF-STEM signal reveals clear changes in the form of upward/downward bumps (Fig.~\ref{FIG4}b,d) at the position of the shear band. According to Eq.\,(2), this indicates a change in density of $(+2.7 \pm 0.4)\,\%$ for the bright and $(-4.3 \pm 0.3)\,\%$ for the dark shear band segment. The shear band width is here about 5\,$nm$. Thus, there are significant differences between in- and ex situ generated shear bands regarding their density, width and shape, which are now discussed in more detail.

First, the shear bands formed during the in situ tensile tests are discussed. The significant features of such shear bands are a spearhead-like shear front, a meniscus-like profile and no apparent shear step. The meniscus is most probably a consequence of necking under the reduced constraint of a 2-D geometry in the thin foil and thus displays a snapshot before failure. Failure occurs when the material is unable to accommodate the accumulated shear strain. Thus, all these features give evidence for catastrophic failure and explain the observed lack of ductility under tension. The spearhead-like shear front implies shear band initiation by structural rejuvenation\cite{shimizu2006yield,greer2013shear,maass2015shear}. Moreover, the observed necking and ripping of shear bands suggests that this is the very beginning of stage two (see third model in the introductory part), where shearing proceeds along the already softened path.

\noindent The lack of distinct alternating density changes in the in situ generated shear bands may have two reasons: First, density changes are not discernible due to an overlap with necking and tapering of the shear bands. Second, the shear bands rip before such features could develop. In support of the latter, we refer to TEM investigations of notched 3-point bending tests, where shear steps had emerged from the tensile side and the corresponding shear bands displayed such characteristic alternating density changes\cite{hilke2019influence,Davani2019}.  

\noindent Moreover, it is worth noting that the actual temperature rise in the shear band cannot be evaluated from the lack of nanocrystallization in the shear bands, since the subsequent cooling process would be too fast to afford immediate nucleation and growth of nanocrystals\cite{matthews2008electron,wilde2011nanocrystallization}.
 
In contrast, shear bands formed upon cold rolling display different features: the shear bands are intact showing no necking, they display distinct alternating density changes along the propagation direction and have considerably narrower widths. As shown in the Supplementary Material (Fig.\,S3), the shear bands exhibited shear steps penetrating through the surfaces with heights between 30\,$nm$ and 200\,$nm$. Maa{\ss} and coworkers reported trends or correlations between the density and both the width and the accumulated shear strain in Zr-based bulk metallic glasses \cite{liu2017shear,maass2020beyond,liu2021strain} and estimated the accumulated shear strain by the ratio of the shear off-set height to shear band width. Applying this procedure to our Al$_{88}$Y$_{7}$Fe$_{5}$ data based on samples deformed by cold-rolling\cite{rosner2014density,schmidt2015quantitative,grove2021plasmon} or high-pressure compression yields shear strain values in the range between 5 and 50 with density changes from  $+6\%$ to $-10\%$ at relative constant shear band widths around $5\pm 1.5$\,$nm$. The presence of intact shear bands (no necking) with emerging shear steps at the surfaces shows that the shear strain has been accommodated successfully. Moreover, the observation that each contrast reversal (density change) is accompanied by a small deflection of the shear band from the main propagation direction (Fig.~\ref{FIG3}) suggests stick-slip motion\cite{klaumunzer2011stick,schmidt2015quantitative}, which also supports the model that nucleation, propagation and arrest of shear bands occur during different stages\cite{cao2009structural,klaumunzer2011stick,greer2013shear,maass2015shear}. 

In summary, we have shown that shear bands generated in situ have a meniscus-like foil thickness reduction, most probably as a consequence of necking under the reduced constraint of a 2-D geometry in the thin foil. With the observation of a spearhead-like shear front, the in situ experiment sheds light on the correlation between the shape and width of shear bands with the stress amplitudes indicating a propagation front mechanism. In comparison with ex situ samples, shear steps accommodating this strain accumulation and distinct alternating contrast changes due to volume/density changes are not observed. An explanation for this behavior may be that in situ generated shear bands rip before such features could develop. Moreover, both in and ex situ experiments suggest that initiation, propagation and arrest of shear bands occur during different stages. 

\section*{Declaration of Competing Interest}
The authors declare that they have no known competing financial interests or personal relationships that could influence the work reported in this paper.

\section*{Acknowledgments}
 This work was partially carried out with support of the Karlsruhe Nano Micro Facility (KNMF, www.knmf.kit.edu), a Helmholtz research infrastructure at Karlsruhe Institute of Technology (KIT, www.kit.edu). We gratefully acknowledge financial support by the DFG via WI 1899/29-1 (Coupling of irreversible plastic rearrangements and heterogeneity of the local structure during deformation of metallic glasses, project number 325408982). The DFG is further acknowledged for funding our TEM equipment via the Major Research Instrumentation Program under INST 211/719-1 FUGG.

\bibliography{LITERATURE-Manuscript-in-situ-SB-final}

\end{document}